\documentclass[12pt,preprint]{aastex}

\usepackage{amssymb}
\usepackage{amsmath}
\shorttitle{Mass segregation in NGC 2298}
\shortauthors{Pasquato et al.}

\begin{document}

\title{Mass Segregation in NGC 2298: limits on the presence of
an Intermediate Mass Black Hole
\footnote{Based on observations made with the NASA/ESA Hubble
Space Telescope, obtained at the Space Telescope Science Institute,
which is operated by the Association of Universities for
Research in Astronomy, Inc., under NASA contract NAS 5-26555.
This paper is associated with program \#11284.}}

\author{Mario Pasquato$^1$, Michele Trenti$^{2}$, Guido De
Marchi$^3$, Michael Gill$^4$, Douglas P. Hamilton$^4$, M. Coleman
Miller$^4$, Massimo Stiavelli$^5$, Roeland P. van der Marel$^5$}

\affil{1 Department of Physics, University of Pisa, Largo Bruno
Pontecorvo 3, I-56127 Pisa, Italy} 

\affil{2
Center for Astrophysics and Space Astronomy, University of Colorado,
Boulder, CO, 80309-0389 USA} 

\affil{3 Space Science Department, European Space Agency, Keplerlaan
1, 2200 AG Noordwijk, Netherlands} 

\affil{4 University of Maryland, Department of Astronomy and Maryland
Astronomy Center for Theory and Computation, College Park, MD
20742-2421 USA }

\affil{5 Space Telescope Science
Institute, 3700 San Martin Drive, Baltimore, MD 21218 USA }

\email{pasquato@df.unipi.it,trenti@colorado.edu}

\begin{abstract}
Theoretical investigations have suggested the presence of Intermediate
Mass Black Holes (IMBHs, with masses in the $100-10000 M_{\sun}$
range) in the cores of some Globular Clusters (GCs). In this paper we
present the first application of a new technique to determine the
presence or absence of a central IMBH in globular clusters that have
reached energy equipartition via two-body relaxation. The method is
based on the measurement of the radial profile for the average mass of
stars in the system, using the fact that a quenching of mass
segregation is expected when an IMBH is present. Here we measure the
radial profile of mass segregation using main-sequence stars for the
globular cluster NGC 2298 from resolved source photometry based on
HST-ACS data. NGC 2298 is one of the smallest galactic globular
clusters, thus not only it is dynamically relaxed but also a single
ACS field of view extends to about twice its half-light radius,
providing optimal radial coverage. The observations are compared to
expectations from direct N-body simulations of the dynamics of star
clusters with and without an IMBH. The mass segregation profile for
NGC 2298 is quantitatively matched to that inferred from simulations
without a central massive object over all the radial range probed by
the observations, that is from the center to about two half-mass
radii. Profiles from simulations containing an IMBH more massive than
$\approx 300-500 M_{\sun}$ (depending on the assumed total mass of NGC
2298) are instead inconsistent with the data at about $3 \sigma$
confidence, irrespective of the IMF and binary fraction chosen for
these runs.  Our finding is consistent with the currently favored
formation scenarios for IMBHs in GCs, which are not likely to apply to
NGC2298 due to its modest total mass.  While providing a null result
in the quest of detecting a central black hole in globular clusters,
the data-model comparison carried out here demonstrates the
feasibility of the method which can also be applied to other globular
clusters with resolved photometry in their cores.

\end{abstract}

\keywords{stellar dynamics --- globular clusters: general --- methods:
  n-body simulations}

\section{Introduction}\label{sec:intro}

Intermediate Mass Black Holes (IMBHs --- with masses of order
${10}^{2} M_\odot$ to ${10}^{4} M_\odot$) have been suggested to form
in the cores of young star clusters \citep[e.g., see][]{por04} and in
dense globular clusters \citep{miller02}. Possible observable features
related to these objects such as shallow cusps in the density and
velocity dispersion profile of the cluster stars were proposed early
on (e.g. see the seminal paper of \citealt{frank_76} and
\citealt{van_der_review,miller04} for a review). Cuspy central Surface
Brightness Profiles (SBPs) have now actually been observed in HST data
on a sizable fraction of GCs \citep[][]{noyola_sbps}, but it is
unclear if they are directly related to IMBHs. Detections based on the
analysis of line-of-sight velocity data have been made but the
evidence does not appear to be conclusive (see the introductory
discussion in \citealt{gill08}). For example, a paradigmatic case is
that of the much debated detection of a $2 \cdot {10}^4 M_\odot$ black
hole in the extragalactic cluster $G1$
\citep[see][]{IMBH_detection_G1, IMBH_undetection_G1,
IMBH_redetection_G1,Trudolyubov,Ulvestad}.

Ideally a direct unambiguous detection of an IMBH is possible in GCs by
measuring orbits of stars bound to the BH, which would also allow a
precise measurement of the central mass. Present-day HST imaging
capability has the accuracy required for this kind of observation but
a significant investment of time spread over multiple epochs is
required. It is therefore necessary to find preliminary criteria to
narrow down the list of candidate GCs for focused follow-up
observations. 

In this paper we apply to NGC 2298 a new method we recently proposed
for assessing the presence of an IMBH \citep[see ][]{gill08}. The idea
is to quantify the amount of mass segregation present in a
collisionally well-relaxed stellar system, that is with a half-light
two-body relaxation time below one billion years. We have shown
through direct N-body simulations that the presence of an IMBH heavily
affects mass segregation of stars in a GC. Systems hosting an IMBH
develop a low degree of mass segregation, as opposed to IMBH-free GCs
in which more massive stars move preferentially towards the center of
the cluster over a relaxation timescale. The presence of the IMBH
tends to equalize the velocity dispersions of all stellar mass
components in the system, thus reducing radial mass segregation.  Note
that the method we propose is not applicable to the most massive
galactic globular clusters such as $G1$ and $\Omega$ Centauri, because
their two-body relaxation time is too long and the amount of observed
mass segregation does not necessarily reflect its long term
equilibrium value. Our investigation thus aims to cover a different
region of galactic GC parameter space compared to the one explored by
current claims of GCs IMBH detection such as
\citet{IMBH_detection_G1} and \citet{noyola_omega}.

The radial profile for mass segregation in main-sequence (hereafter
MS) stars can be readily measured in GCs via observations with
sufficient angular resolution to resolve individual sources in the
crowded cores of these systems \citep[e.g. see][]{demarchi07}. In this
paper we use archival HST ACS observations of NGC 2298 as analyzed by
\citet{demarchi07}. These data are ideal to measure the radial
variations of mass segregation because they range from the center out
to more than twice the half-light radius. The observed radial
variation in the mean MS mass is then compared to expectations based
on the numerical simulations of \citet{gill08}, which allow us to
constrain the mass of a central BH in the system.

This paper is organized as follows. In Section~\ref{sec:obs} we
discuss the HST observations for NGC 2298 and our data analysis, which
is then compared to our N-body simulations in Section~\ref{sec:ns}. We
summarize and conclude in Section~\ref{sec:con}.

\section{Observations and Data Analysis}\label{sec:obs}

We study the mass segregation of MS stars of NGC $2298$ using deep
HST/ACS observations in the F606W and F814W bands. The field covers an
area of $3.4 \cdot 3.4$ arcmin$^2$ around the cluster center,
extending to more than twice the cluster's half-light radius
\citep[see ][]{harris}. The data reduction is described in
\citet{demarchi07}, who derive the color-magnitude diagram that we use
in our analysis. The data have $10 \sigma$ detection limits of
$m_{F606W} \approx 26.5$ and $m_{F814W} \approx 25$ with a
completeness at the detection limit above $50\%$. Background sources
contamination is essentially negligible for this cluster
(\citealt{demarchi07}). We assume a distance modulus of $15.15$ mag
(distance of about $12.6$ kpc), and color excess $E(B-V) = 0.14$
from \cite{harris}. From the color magnitude diagram we infer the mass
of each individual main sequence source using the mass-luminosity
relation from \citet{baraffe_MLrelat} and assuming a metallicity
$[Fe/H] = -1.85$ for the cluster \citep{harris}. Our MS stars catalog
consists of objects with masses in the range $[0.2:0.8] M_{\sun}$,
where the lower limit is set by completeness cutoff and the upper
limit by the turn-off mass.

\subsection{Cluster Properties}\label{sec:models}

The Surface Brightness Profile for NGC 2298 is part of the
\cite{cat_trager} compilation of photometric data on galactic
GCs. \cite{cat_mclaugh} fitted a single-mass King model to it,
obtaining a total mass of $3.09 \cdot 10^{4} M_\odot$ while
\citet{demarchi07} derive a total mass of $5 \cdot 10^{4} M_\odot$
based on a multi-mass dynamic model. The \cite{harris} catalog reports
a projected half-light radius $r_{hl}$ of $46.8$ arcsec, while
\citet{cat_mclaugh} derive $r_{hl}= 45.4$ arcsec, corresponding to
$2.35$ parsecs. \citet{demarchi07} find instead a larger half-mass
radius --- that is $r_{hm} = 72$ arcsec --- from their multi-mass
model.

From the structural parameters of NGC2298 we can derive its half-light
relaxation time, defined in physical units as \citep{djo93}:
\begin{equation}\label{eq:trel}
t_{rh} = \frac{8.9 \cdot 10^5 yr}{\log(0.4N)} \times \left (
 \frac{1 M_{\sun}}{\langle m_*\rangle} \right )\times \left( \frac{M_{tot}}{ 1 M_{\sun}} \right )^{0.5} \times 
\left( \frac{r_{hl}}{ 1 pc} \right )^{1.5},
\end{equation}
where $\langle m_* \rangle = M_{tot}/N$ is the average mass of a star
(including dark remnants), $N$ is the number of stars and $M_{tot}$ is
the total mass of the system. 

Assuming an average stellar mass of $0.5 \mathrm{M_{\sun}}$, the
half-light relaxation time for NGC 2298 is ${10}^{8.41}$ yr using the
\citet[][]{cat_mclaugh} structural parameters. If we consider instead
the \citet{demarchi07} modeling \footnote{Note that the relaxation time of 3.1 Gyr quoted
 in \citet{demarchi07} for NGC 2298 is the result of a typo in the
 output of the code they used to contruct the dynamic model.},
Eq.~\ref{eq:trel} yields $t_{rh} = 10^{8.76}$ yr. In both cases
NGC2298 appears to be a dynamically old cluster even after accounting
for a significant mass loss in the cluster during its evolution, as
suggested by \citet{baumgardt08}: if the cluster was originally four
times as massive and twice as large
\footnote{\citet{trenti_heggie_hut} find that the half-mass radius
  stays approximately constant while a star cluster is being tidally
  disrupted} there would still have been enough time to relax, as the
initial relaxation time would have been $t_{rh}(t=0) \sim 3 Gyr$.

Further evidence for an old dynamic age comes from the shape of the
mass function of NGC $2298$, which shows a depletion of low mass
stars. This makes the cluster a good candidate for our search because
it would be expected to have undergone core collapse, unless there is
a central source of energy capable of halting the core contraction,
such as an IMBH or a significant population of primordial binaries
\citep[e.g. see ][]{trenti_heggie_hut,trentiea07b}. While the current
formation scenarios for IMBHs in GCs assume a higher cluster mass than
the current NGC2298 mass, its probable initial mass --- up to $1.4
\times 10^5 \mathrm{M_{\sun}}$ \citep{baumgardt08} --- was well in the
range required for IMBH formation.

In order to consistently compare the data with our sample of numerical
simulations, we identify the projected \emph{half-mass radius in main
sequence stars within the ACS field of view} --- which we call
$r_{hm}$. Given that the mass-to-light ratio of globular clusters has
radial gradients, this quantity is different from the standard
half-light radius used in the literature. To measure $r_{hm}$ we bin
the star counts in cluster-centric radius and sum over star mass,
applying the completeness correction appropriate for the given
cluster-centric radius and star magnitude. We obtain the surface
density profile of the cluster from main sequence stars and then apply
the non-parametric spline-smoothing technique described in
\cite{mypaper} to the profile obtaining our best estimate of the total
main sequence mass of the cluster in the field of view and of the
respective half-mass radius. We do not
extrapolate the light profile outside the ACS field which might lead
to the discrepancies in the determination of the structure of the
cluster discussed above, but rather include the effect of a finite
field of view into the simulation analysis (see
Section~\ref{sec:ns}). We derive $r_{hm}=49.0$ arcsec, slightly larger
than $r_{hl}$ reported in \cite{harris}. This is not surprising, as
the light profile tends to be dominated by red giant stars, more
massive on average than MS stars and therefore more centrally
segregated \citep[for example, see][]{hurley2007}.

\subsection{Mass Segregation}\label{sec:mass_segr}

We quantify mass segregation through the radial variation of the mean
mass ${\langle m \rangle}_{MS}$ of MS stars. At the center of the
system, where high mass stars segregate through energy equipartition,
the mean mass of stars in such range is expected to be higher than in
the outer region of the system \citep{spitzer1987_book}. To construct
${\langle m \rangle}_{MS}(r)$ we apply the completeness correction as
follows:
\begin{equation}
 \label{UnbinnedCompleteness}
 \langle m \rangle_{MS} (r) = \frac{\sum_i M_i/f(m_i,
 r_i)}{\sum_i 1/f(m_i, r_i)}
\end{equation}
where $M_i$ is the star's mass and $f(m,r)$ is the completeness for a
star of magnitude $m$ at cluster-centric 2D radius $r$. To apply the
completeness correction, we assumed a one to one map from mass to
luminosity ($M_i$ to $m_i$) for main-sequence stars
\citep[specifically that of ][]{baraffe_MLrelat}. The color
information is of course used to select main sequence stars. The sum
over the index $i$ is carried out over all stars in an annulus around
the radius $r$. Since the completeness function was defined in both
radial and magnitude bins by \citet{demarchi07}, we define a
continuous $f(m, r)$ by bilinear interpolation. As a compromise
between obtaining high spatial resolution and minimizing Poisson
fluctuations we calculate ${\langle m \rangle}_{MS}(r)$ using $20$
concentric annuli comprising each $5\%$ of the stars in number.

Errors on each radial annulus are obtained using a bootstrap technique
\citep{bootstrap}. We proceed as follows: Let $\{M_i\}_{i=1,N}$ be
the catalog of observed main sequence star masses within a radial
annulus. We use this catalog to generate 100 synthetic catalogs. To
construct a synthetic catalog, we first extract $N$ random numbers
$\{q_j \}_{j=1,N}$ from a uniform random number generator in the range
$[1:N]$. The synthetic catalog of masses is then defined as
$\{M_{q_j}\}_{j=1,N}$, that is we extracted with replacement and
uniform probability from the original
data-set. Eq.~\ref{UnbinnedCompleteness} is then applied to each
synthetic catalog of a radial annulus and the 1 $\sigma$ error on the
observed ${\langle m \rangle}_{MS}(r)$ is defined as the standard
deviation of the sample of the 100 synthetic catalogs for that radial
position.

The observed mass segregation profile, normalized to the average mass
measured around $r_{hm}$ (using the average MS mass measured between
0.8 and 1.2 $r_{hm}$) is indicated as $\Delta m (r)$ and shown in
Fig. \ref{fig:mainplot_type1}. From the plot it is clear that the
cluster has a marked radial variation in the MS mass, with a
difference of about $0.14 M_{\sun}$ from the center to the outermost
region at $r > 2 r_{hm}$. This is about $30\%$ of the average mean MS
mass of the system $0.529 \pm 0.002 M_{\sun}$.

\section{Comparison with Numerical Simulations}\label{sec:ns}

The numerical simulations used in this paper have been carried out
with a state-of-the-art direct N-body code for star cluster dynamics,
NBODY6 \citep{aar03}. NBODY6 has been modified as discussed in
\citet{trentiea07b} to improve accuracy in the presence of an IMBH,
and uses regularization of close gravitational encounters without any
softening, guaranteeing high accuracy on the integration of the star
trajectories even in the proximity of the black hole.
 
The details of the N-body simulations --- along with a mass
segregation analysis --- are discussed in \citet{gill08}, here we
summarize their properties (see also Tab. \ref{tab:sim}).  We use from
$16384$ to $32768$ particles. Their individual masses are drawn from
an initial mass function (IMF, either \citealt{salpeter} or
\citealt{millerscalo}) and then evolved through an instantaneous step
of stellar evolution to $\approx 12$ Gyr of age using the
\citet{hurley2000} tracks. Our stellar mass black holes are in the
$[5:10] \mathrm{M_{\sun}}$ range. The stellar evolution step evolves
the cluster IMF to match the turn-off mass of $\approx 0.8 M_{\sun}$
observed in NGC 2298. Simulations also have up to $10\%$ of primordial
binaries and include the effects of a galactic tidal field.

Simulations are carried out until complete tidal disruption of the
systems, which happens after about 20 initial half-light relaxation
times $t_{rh}$ (defined in Eq.~\ref{eq:trel}), a timescale longer than
the age of the Universe for a typical globular cluster. As discussed
in \citet{gill08}, the simulations settle down in a quasi-equilibrium
configuration with respect to mass segregation after a few relaxation
times. At this point, there are three regimes for the asymptotic value
of mass segregation measured as the difference in the main sequence
mass between the center and the half mass radius\footnote{Note that
the quantity $\Delta \langle m \rangle$ defined in \citet{gill08} has
a slightly different normalization with respect to $\Delta m (r=0)$
used in this work. We define $\Delta m (r=0)$ as the difference in
mean MS mass between the center and mean MS mass in the radial annulus
$r/r_{hm} \in [0.8, 1.2]$. \citet{gill08} use instead a smaller
annulus around the half mass radius, defined as that containing 5\% of
mass of the cluster and centered on $r_{hm}$. In Table~1 we report
both indicators for comparison.} ($\Delta \langle m \rangle$): a low
value ($\Delta \langle m \rangle < 0.07$) corresponding to an high
probability of harboring an IMBH, a high value ($\Delta \langle m
\rangle> 0.1$) associated to a low probability of harboring an IMBH
and an intermediate regime, where models with and without an IMBH both
can be found. In addition, the mass segregation profile measured
through $\Delta m (r)$ and/or $\Delta \langle m \rangle$ is a
differential measure normalized to the half-mass radius, so the
average value of the MS star mass over the whole cluster is not
important and runs with either a \citealt{salpeter} or
\citealt{millerscalo} IMF have similar $\Delta m (r)$ profiles.

Furthermore, our N=32k runs starting with a \citet{millerscalo} mass
function have a similar shape of the mass function in main sequence
once they have lost about 75\% of their initial mass (as estimated for
NGC 2298 by \citealt{baumgardt08}). This is shown in the left panel of
Fig.~\ref{fig:mass_function} for our simulation 32km at $t = 16
t_{rh}$: dynamic evolution of the cluster and the subsequent
preferential ejection of light stars has naturally evolved the initial
Miller \& Scalo distribution to match that observed in NGC 2298,
especially for the simulation with no BH. To ensure a self-consistent
comparison of the observations with our simulation,
Fig.~\ref{fig:mass_function} has been obtained by taking into account
the completeness limit of the data when ``observing'' the simulation:
we have projected the simulation snapshots along random lines of
sight, then rejected with probability $1-f(m,r)$ particles with
magnitude $m$ at a 2D radius $r$.

To construct the mass segregation profile of our simulations, we use
all the snapshots from \citet{gill08} with $N \geq 16384$
particles. We restrict the analysis to the time interval between $7$
and $9 t_{rh}$, a period in which the simulated clusters have reached
their asymptotic amount of mass segregation but still have more than
half of the initial number of particles (except for the plot shown in
Fig.~\ref{fig:mass_function}, obtained using snapshots between 15.5 and
16.5 $t_{rh}$). This gives us $26$ snapshots per simulation for runs
with $N=16384$, and $32$ snapshots for runs with $N=32768$ for a total
of $324$ snapshots available for comparison with observations. As in
this paper we aim at analyzing the simulations as similarly as
possible to the NGC 2298 data, we recast the analysis in terms of the
full radial profile for mass segregation rather than limiting to
$\Delta \langle m \rangle$ as in \citet{gill08}. Only line-of-sight
projected quantities are used and only main sequence stars are
considered in the analysis. For those runs with primordial binaries,
binary stars are projected onto the main sequence for single stars and
treated as a single star with mass equal to that of the heavier main
sequence member (see \citealt{gill08}). The analysis is also
restricted to MS stars more massive than $0.2 M_{\sun}$ to match the
completeness limit of the observations. Under these assumptions we
calculate the center of visible projected mass for each simulation
snapshot (this turns out not to differ significantly from the center
of mass of all gravitationally bound particles) and a first guess of
$r_{hm}$. We then repeat our analysis including only particles lying
within $2 r_{hm}$ from the center and recompute both the center of
visible mass and the final $r_{hm}$ value. This procedure closely
resembles the observational limit on the field of view of our data,
which extends to about $2 r_{hm}$. To bring the number of particles
``observed'' in a simulation snapshot close to the actual number of
stars in NGC 2298 we sum three independent projections for each
snapshot. In this way we almost reach a 1:1 ratio of particles to
stars for our larger $N=32768$ runs.

Fig. \ref{fig:mainplot_type1} shows the results from the
\citet{gill08} sample of snapshots superimposed to the observed mass
segregation profile derived in Section ~\ref{sec:obs}. The blue shaded
area corresponds to the region defined by the $2 \sigma$ contours for
points derived from all the snapshots of all runs with an IMBH (thus
encompassing simulations with different IMF and binary fractions as
reported in Table~1), while the green shaded area is the $2 \sigma$
region for runs with no central BH (including all the no BH runs in
Table~1).

The data-points for NGC 2298 are fully encompassed by profiles derived
by simulations run without a black hole and the central profile
(innermost two measurements) is outside the region associated to
simulations with a central BH. The $1\sigma$ error bars associated to
the observed points are smaller than the scatter from simulation
snapshots, thus the main factor in setting the confidence level at
which we can exclude a central black-hole is set by the variance in
the simulations. The upper right inset of
Fig.~\ref{fig:mainplot_type1} shows the upper envelope of all $162$
mass segregation profiles obtained from snapshots
of simulations containing an IMBH.  The two central observed data
points lie above such envelope at a combined confidence level greater
than $2 \sigma$ confidence level, providing us with a quite stringent
test against the possibility that the observed high amount of mass
segregation in NGC 2298 is the result of a random fluctuation of a
system with a central IMBH. If this were the case, two $> 2 \sigma$
fluctuations both in the simulations and in the observations would be
required, thus we can exclude this scenario at about $3 \sigma$ level.

To better quantify the influence of different initial conditions on
the expected variance of the mass segregation profile, we used a
second, different approach. Rather than calculating the variance of
all sample of simulations with or without a central IMBH, we compute
the variance in $\Delta m (r)$ separately for each run. In
Fig. \ref{fig:mainplot_type2} we then define the blue and green shaded
areas as the envelope of the $2 \sigma$ regions of each run. In this
way simulations with different IMF and binary fractions are treated
separately and the systematic differences between them are not treated
as random error. The two shaded areas are therefore larger by
construction than in Fig. \ref{fig:mainplot_type1}. Still a clear
separation is present between runs with and without an IMBH. showing
that systematics due to different IMF and binary fractions are not a
concern for our results.

Fig.~\ref{fig:mainplot_type2} is also based on four additional runs
with respect to the \cite{gill08} runs that were used to construct
Fig.~\ref{fig:mainplot_type1}. These additional runs, listed at the
end of Table~1, have been generated using the same initial conditions
of run 16kmbh in \citet{gill08}, except for a different seed of the
random number generator. These runs have been added to further study
mass segregation in presence of an IMBH when there is a Miller \&
Scalo IMF. Note that run 16kmbh in \citet{gill08} is the one which
shows the highest degree of mass segregation among those with an IMBH,
even though both its $N=8k$ and $N=32k$ counterparts in \citet{gill08}
have a lower degree mass segregation. Our additional $N=16k$ runs show
on average a slightly lower degree of mass segregation as reported in
Table~1 (eight column), even though one of the four runs is very close
to the original \citet{gill08} results. In preparing
Fig.~\ref{fig:mainplot_type2} we treated all these five 16kmbh runs
independently, thus the $2 \sigma$ envelope shown in the figure
includes a $>99\%$ confidence level area with respect to this specific
initial condition.  Note also that the addition of 106 snapshots from four new runs have
naturally lead to a slight increase of the \emph{maximum} amount of mass
segregation measured in the simulations (plotted in the upper right
inset of Fig.~\ref{fig:mainplot_type2}) compared to the inset of
Fig.~\ref{fig:mainplot_type1}.  But these additional snapshots also increase the confidence level for BH rejection:  the inset in
Fig.~\ref{fig:mainplot_type2} is based on a total of $266$ snapshots and still
none of those snapshots reaches the data whithin the observed error
bar (in particular with respect to the second innermost data point).

If we restrict our analysis to snapshots that have a global mass
function similar to that observed in NGC 2298, our results are
stronger. The right panel of Fig.~\ref{fig:mass_function} shows a
comparison of the observed $\Delta M(r)$ with the expectations from
our N=32k Miller\&Scalo simulations evolved around $t=16 t_{rh}$, when
$N\sim 8000$ particles remains in the system (a mass loss consistent
with the one estimated for NGC 2298 by \citealt{baumgardt08}). The
global mass function of the simulations at that point is an excellent
match to the observed one, especially for the run without central
BH. Evaporation of a significant fraction of stars has marginally
lowered the amount of central mass segregation compared to that
measured at $t \approx 8 t_{rh}$ for the same runs. The contour areas
for the simulated $\Delta M(r)$ in this figure are contained within
those of Fig.~\ref{fig:mainplot_type1}, but are significantly more
compact in the central region. Therefore from
Fig.~\ref{fig:mass_function} the presence of a central BH can be
rejected at a significance level much higher than $3\sigma$. In
addition, we get a fully consistent representation of the mass
segregation profile from the simulations without a central BH. The
quantitative match of the observed mass segregation profile to the
numerical expectations also provide a posteriori evidence that NGC
2298 has reached its equilibrium value for mass segregation and thus
it is collisionally relaxed (see discussion in
section~\ref{sec:models}).

All our simulations that include an IMBH have a ratio of about $1\%$
between the total cluster mass and the black hole mass. Assuming a
total mass for NGC 2298 of $3.09 \times 10^4 \mathrm{M_{\sun}}$, this
implies that we can exclude the presence of a central BH of mass
$\gtrsim 300 \mathrm{M_{\sun}}$. The latter limit would increase to
$500 \mathrm{M_{\sun}}$ if the total mass of NGC2298 is instead
$\approx 5 \times 10^4 \mathrm{M_{\sun}}$.

\subsection{Systematic Uncertainties}\label{sec:syst}

In order to compare simulations to observations there is need to fix a
radial scaling. We choose to normalize the mass segregation profile
$\Delta m (r)$ to its value at $r_{hm}$, as discussed in
Section~\ref{sec:obs} and shown in Fig. \ref{fig:mainplot_type1}. A
correct measurement of $r_{hm}$ is therefore vital for a consistent
comparison between simulations and observations. For the simulations,
we have a complete control over the observables in the system, but we
have to evaluate possible biases in the measure of $r_{hm}$ from the
photometry of NGC 2298. Mis-estimating the value of mean MS mass at
$r_{hm}$ due either to local fluctuations of ${\langle m
\rangle}_{MS}(r_{hm})$ or to a wrong estimate of $r_{hm}$ can then in
principle lead to an un-physical shift of the whole profile upwards or
downwards along the vertical axis.

Local fluctuations in the average main sequence mass are reduced by
our choice to normalize the $\Delta m (r)$ profile using the mean MS
mass between $0.8 r_{hm}$ and $1.2 r_{hm}$, instead of the punctual value
at exactly $r_{hm}$ (see Section~\ref{sec:obs}). We note in passing
that also choosing a smaller interval ($[0.9, 1.1] r_{hm}$) yields
comparable results, although with more noise.

To assess the impact of a mis-determination of $r_{hm}$, we repeat our
analysis using values for $r_{hm}$ shifted by $\pm 4$ arcsec from our
best estimate $r_{hm}=49$. A change in $r_{hm}$ of this amount does
not critically affect our conclusion, as it shifts $\Delta m$ by less
than the $1 \sigma$ error associated to the measure. The amount of
observed mass segregation is decreased if one adopts a smaller value
for $r_{hm}$.  Only a significant change in $r_{hm}$ ($r_{hm} \lesssim
40$ arcsec) would move the measurement within the amount of
segregation typically associated to the presence on an IMBH.

Another source of systematic uncertainty that needs to be addressed is
the determination of the center of the cluster.  \cite{noyola_sbps}
point out that center determination of GCs is a difficult problem,
with literature/ground based coordinates of GC centers sometimes being
inaccurate. Miscentering of the cluster in the analysis of
observational data could in principle lead to an artificially
shallower mass segregation profile in the central part of the
cluster. If this is the case, then the confidence level of our null
result would just increase.  However, miscentering is not likely to be
a significant issue in our case. Not only did we use high resolution
data from HST but we determined the center based on the mass of main
sequence stars, not on the total light from the cluster. Therefore the
Poisson fluctuations in the small number of red giant stars that
might dominate the light profile of a globular cluster do not affect
significantly our analysis. To quantify the random error on our adopted
center position we carried out a Monte Carlo bootstrap resampling test.
We generated $100$ synthetic samples of stars by extracting with
replacement main sequence stars from the observed catalog. We then
re-computed the center of each synthetic sample and used the $100$
center coordinates thus obtained to calculate the $1 \sigma$
fluctuation of the center position. The value we recover from our
Monte Carlo test is below $0.4$ arcsec.

\subsection{Applicability of mass segregation indicator to constrain IMBH presence}

In \citet{gill08} we suggest two conservative criteria to ensure that
a stellar system has reached its long-term mass segregation profile:
(1) have a tidal to half-light radius $r_t/r_h \gtrsim 10$ and (2)
have an half-light relaxation time below $1$ Gyr. NGC 2298 fulfills
the second condition, even assuming a worst case estimation of its
structural parameters (see Section~\ref{sec:models}), but marginally
fails the first one, as $r_t/r_h = 8.3$. As discussed in
\citet{gill08}, these requirements are sufficient, but not necessary
for the mass segregation to have reached its equilibrium value. Based
on a detailed dynamic modeling of the cluster in
Section~\ref{sec:models} we demonstrate that the cluster appears to be
well relaxed and its stars to have reached energy equipartition,
despite its significant mass loss due to tidal evaporation.
Furthermore, if there is some primordial mass segregation, like it has
been suggested for NGC 2298 by \citet{baumgardt08}, then the time
needed to develop the asymptotic mass segregation profile becomes
shorter. This provides a further validation for the application of our
analysis to this system.

\section{Discussions and Conclusions}\label{sec:con}

\begin{figure}
\plotone{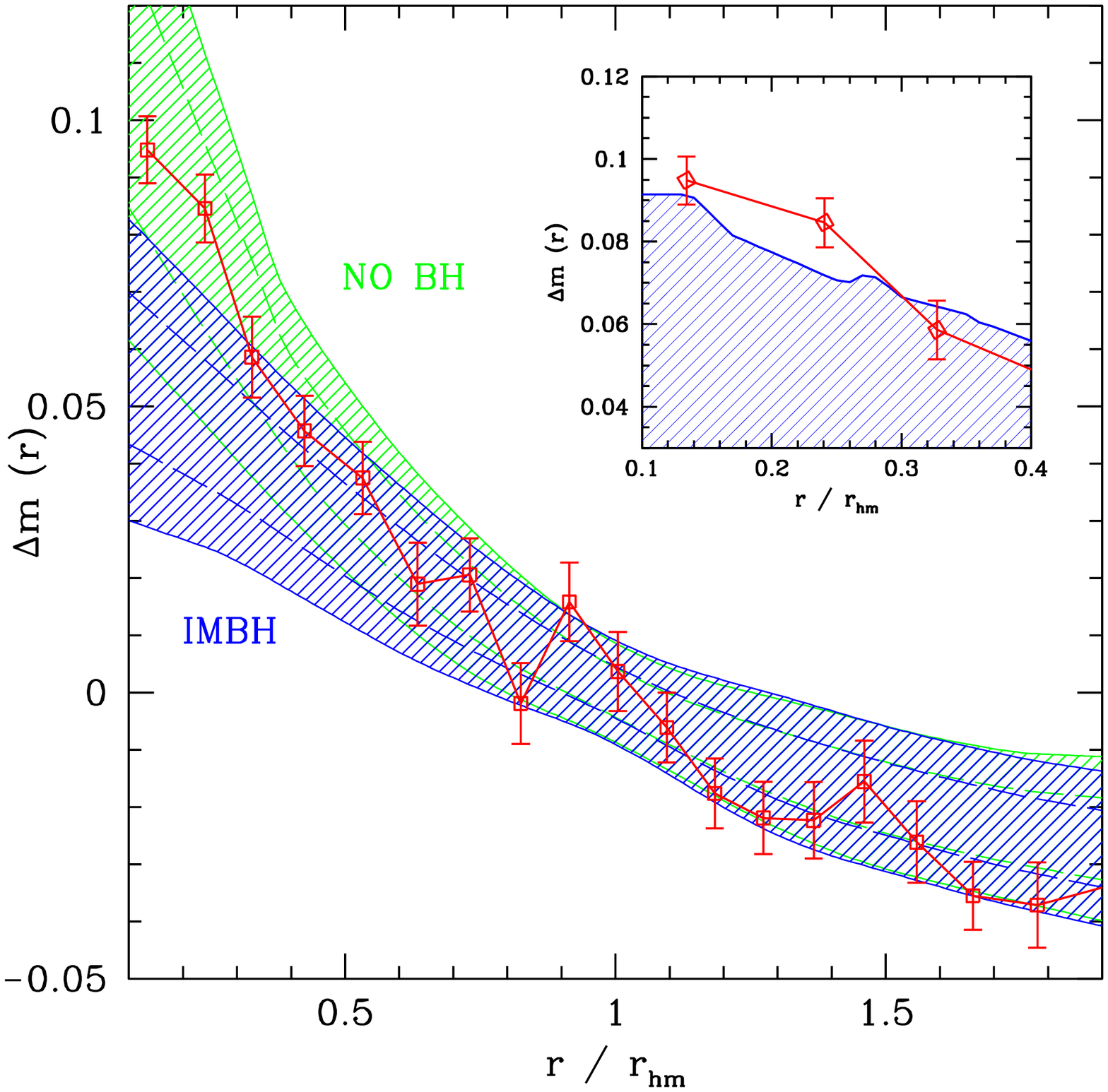}
\caption{Observed radial mass segregation profile ($\Delta m (r)$
measured in $M_{\sun}$) for NGC 2298 (red points with $1 \sigma$ error
bars), compared to expectations from numerical simulations. In the
main panel the blue (IMBH) and green (NO BH) shaded areas represent
the $2 \sigma$ confidence level area for the profiles from the
ensemble of snapshots with $N \geq 16384$ published in \citet{gill08}
(see Table~1 for a description of these runs).  Also shown as long
dashed lines are the inner $1 \sigma$ regions.
The small inset shows the inner observed data
points compared against the upper envelope of all the profiles
associated to snapshots with a central IMBH, that is against the
maximum mass segregation measured in the simulations with an
IMBH. Mass segregation in NGC 2298 appears typical for a system
without a central BH.}
\label{fig:mainplot_type1}
\end{figure}

\begin{figure}
\plottwo{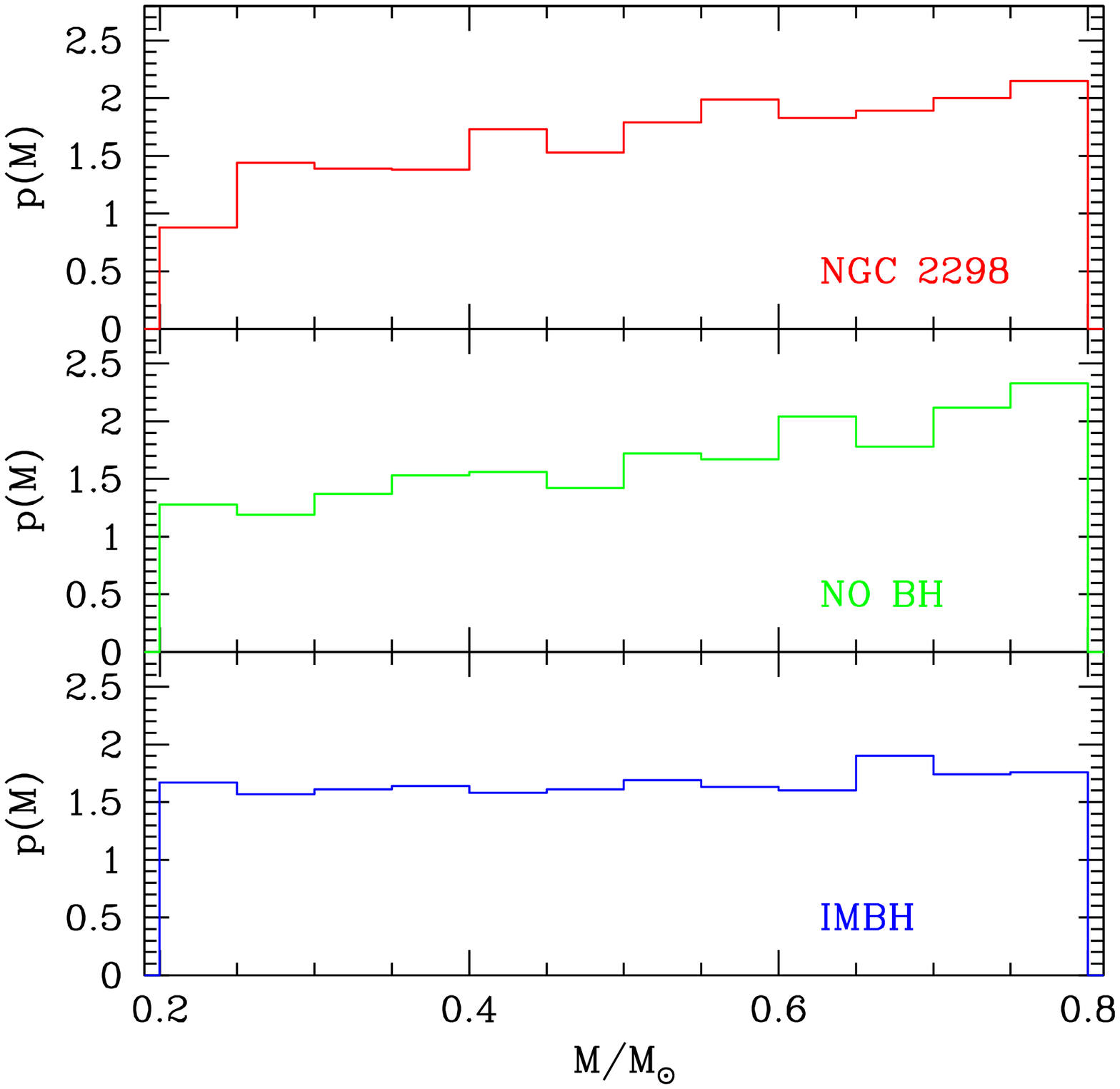}{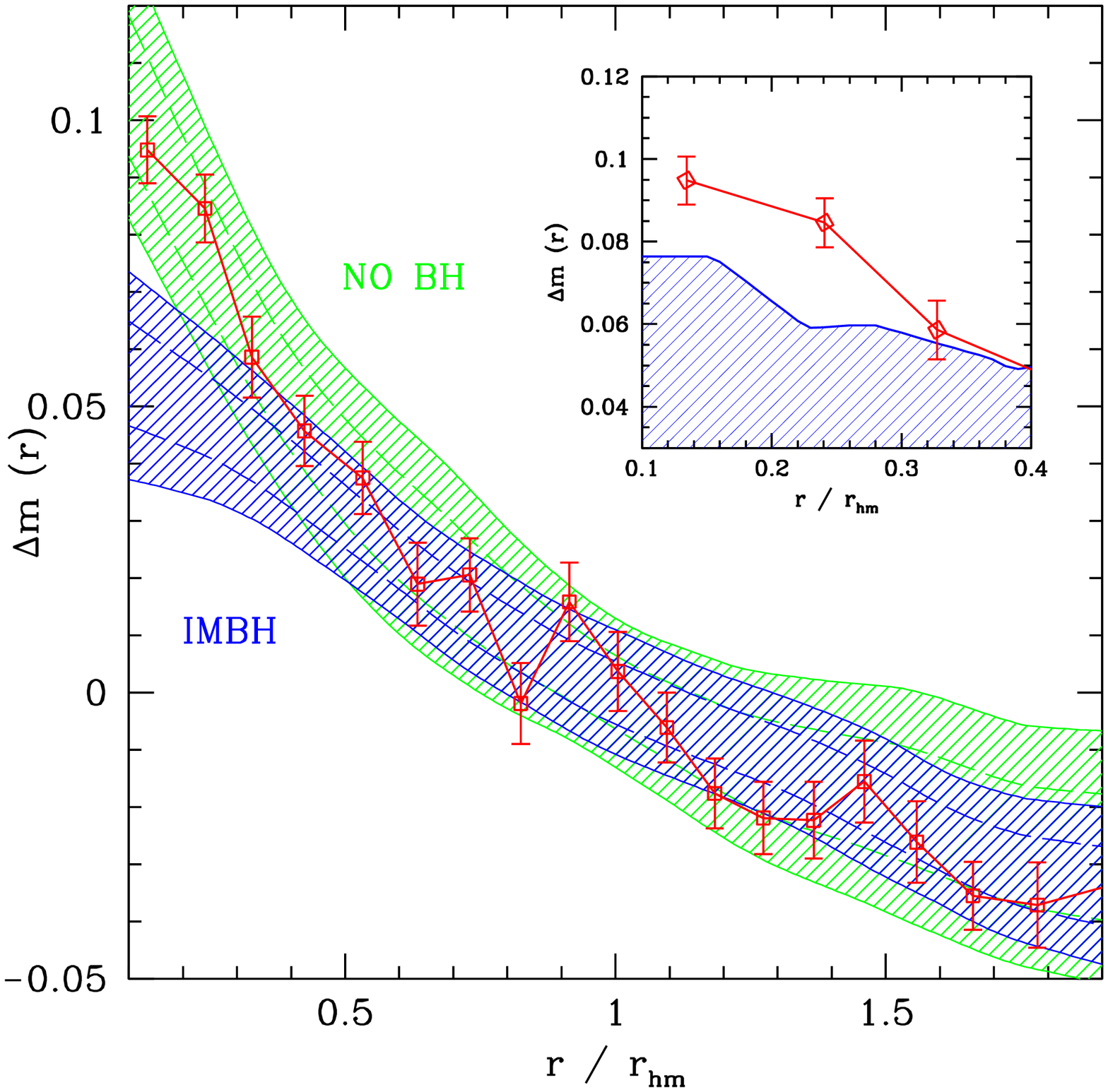}
\caption{Left panel: Observed main-sequence mass function for NGC 2298
(red line) compared to that of our 32km and 32kmbh (NO BH: green line,
IMBH: blue line) simulations at time $t=16t_{rh}$ when about $75$ \%
of the initial mass is lost, as estimated for NGC 2298 by
\citet{baumgardt08}. The simulations started with $N=32768$ particles
from a \cite{millerscalo} IMF and have been projected in 2D and
``observed'' with a field of view extending to $2 r_{hm}$ assuming the
completeness of the NGC 2298 data (see section~\ref{sec:ns}). The IMBH
simulation has a mass function less depleted of light stars because of
its reduced mass segregation. Right panel: Observed radial mass
segregation profile ($\Delta m (r)$) like in
Fig.~\ref{fig:mainplot_type1} but with theoretical expectations based
only on the N=32k snapshots taken in the time interval $t \in
[15.5:16.5]~t_{rh}$. }
\label{fig:mass_function}
\end{figure}

We have analyzed ACS-HST data covering the globular cluster NGC 2298
to quantify the radial variation of mass segregation, measured from
main sequence stars in the range $[0.2:0.8] M_{\sun}$. For dynamically
relaxed systems such as NGC 2298 (that is with a half-mass two-body
relaxation time well below the Hubble time), the degree of mass
segregation present can shed light on the presence of a central
intermediate mass black hole \citep[see][]{gill08}. The observed mass
segregation profile (see Fig.~\ref{fig:mainplot_type1}) has been
compared to the expectation from a sample of direct N-body simulations
with and without a central IMBH. The simulations have been analyzed as
closely as possible to the observational data. We find that NGC 2298
exhibits a fair amount of mass segregation, with the average main
sequence stellar mass being larger at the center compared to that at
the half-mass radius by $\Delta m (r=0) \approx 0.095
\mathrm{M_{\sun}}$. This observed radial variation of the main
sequence mass is fully encompassed by curves derived from simulations
without a black hole. None of the simulation snapshots with a central
IMBH presents such a large degree of mass segregation and the two
innermost data-points lie each at $2 \sigma$ above the maximum mass
segregation measured in the simulations, for a nominal combined
confidence level of about $3 \sigma$. At this confidence level we can
thus consider unlikely the presence of an IMBH of mass $M_{BH} \gtrsim
300 - 500 \mathrm{M_{\sun}}$ (depending on the total mass of the
system). Certainly NGC 2298 does not appear a promising candidate for
follow-up observations to search for an IMBH such as with a proper
motion study. This could partly have been expected based on the small
size of NGC 2298, because the proposed scenarios for IMBH formation in
GCs require a higher cluster mass \citep{miller02,por04} than its
present-day mass. On the other hand, dynamical interaction with the
galactic environment is likely to have stripped the majority of the
cluster's initial mass \citep{baumgardt08}, which could have been well
above the threshold for runaway collapse of massive stars.

A critical assumption in our analysis is a proper evaluation of the
current dynamic state of NGC2298 and in particular of its half-light
relaxation time, which is complicated by the presence of potentially
inconsistent structural parameters in the literature (see our
discussion in Section~\ref{sec:models}). Even assuming a worst case
scenario with respect to the structural parameters published, the
\emph{current} half-light relaxation time we derive is below $\sim 0.6
Gyr$, guaranteeing that the cluster had enough time to reach energy
equipartition and its equilibrium value of mass segregation, even if
it has lost most of its initial mass and thus its \emph{initial}
half-light relaxation time could have been as high as $\sim 3$
Gyr. With respect to this issue, note that \citet{gill08} recommend to
apply the mass segregation analysis to well relaxed clusters, defined
as having a \emph{current} $t_{rh} \lesssim 1Gyr$ and $r_t/r_{hl}
\gtrsim 10$ in order to avoid false positives, that is clusters that
have a low amount of mass segregation not because there is an IMBH but
because the cluster has not yet fully developed mass segregation. In
the case of NGC2289 we are in the opposite regime as we find more mass
segregation than expected in presence of an IMBH. Furthermore, we have
a quantitative match of the observed radial profile of mass
segregation with the expectations from our numerical models. This is a
further evidence that NGC2298 is collisionally relaxed and our
analysis of the full radial profile of mass segregation goes beyond
the single point measure discussed in \citet{gill08}.

The absence of a central IMBH in NGC2298 has an interesting
consequence for the dynamics of the system. The cluster has a rather
large core and does not appear to have undergone core collapse despite
its advanced dynamic age \citep{collapsed_yet}. This means that there
must be another source in the core capable of generating kinetic
energy through gravitational encounters, such as a population of
primordial binaries with a number density $\gtrsim 0.05$
\citep[e.g. see][]{heggie_trenti_hut,trenti_heggie_hut}. The
color-magnitude diagram for the cluster has indeed a widened main
sequence which allows us to get a lower limit of $0.04$ for the binary
fraction in the core and a likely fraction two to three times higher.

\begin{figure}
\plotone{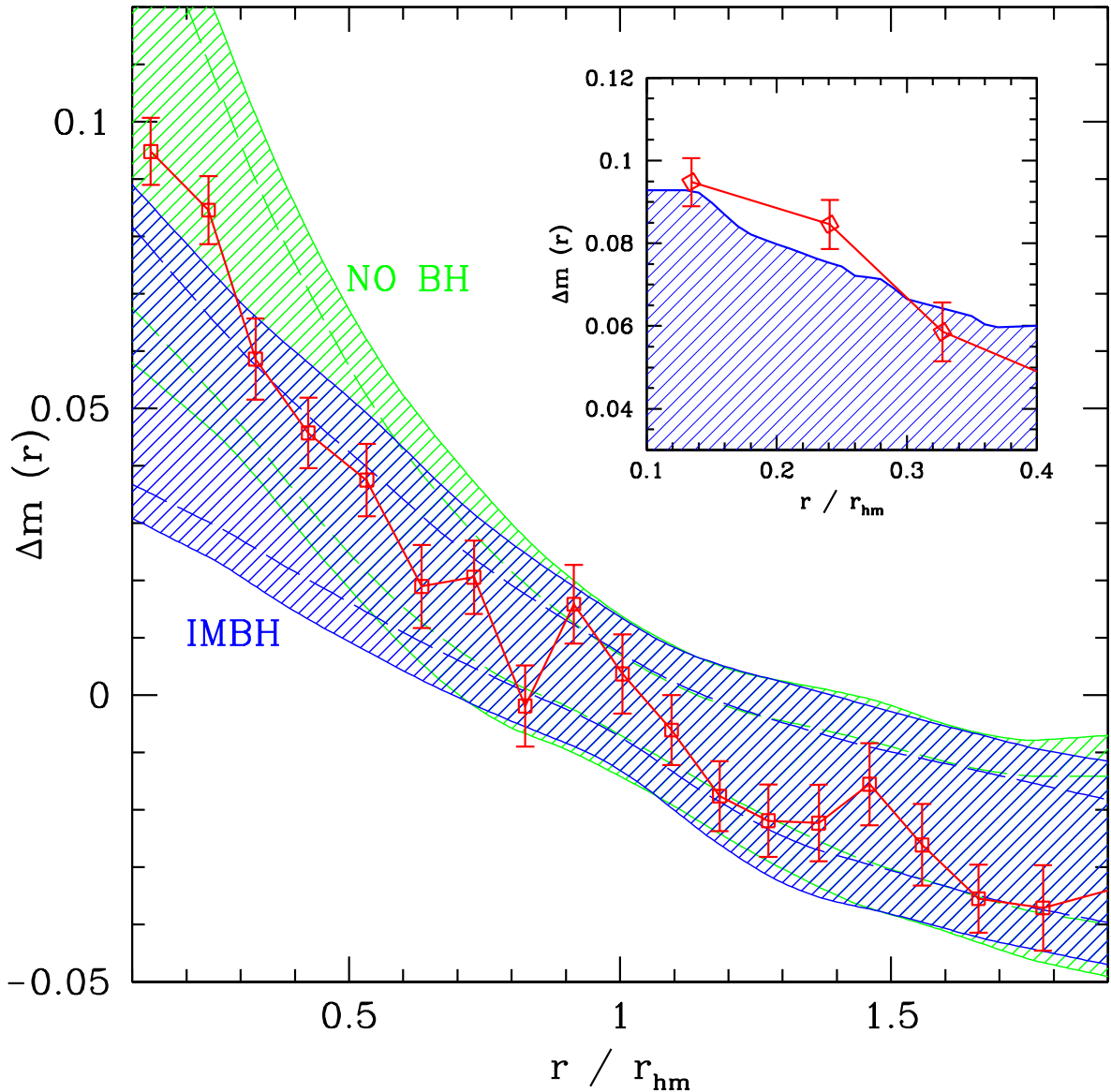}
\caption{Observed radial mass segregation profile for NGC 2298, as in
Fig. \ref{fig:mainplot_type1}. The blue (IMBH) and green (NO BH)
shaded areas are constructed here using a different approach compared
to Fig. \ref{fig:mainplot_type1}: (i) we include four additional runs
with a central IMBH (whose initial conditions lead in \citealt{gill08}
to the highest amount of observed mass segregation; see detail in
Section.~\ref{sec:ns}); (ii) for each simulation run we first define
its $2 \sigma$ confidence level area for mass segregation and then
construct the global $2 \sigma$ areas for the ensembles with and
without a central BH taking the envelope of the individual runs
confidence regions. This procedure better highlights run to run
variations in mass segregation associated to different initial
conditions and by construction results in larger uncertainty
regions. It is therefore an extremely conservative approach and
represents a stronger test than Fig. \ref{fig:mainplot_type1} to
reject the presence of a central BH. The upper small inset is defined
as in Fig. \ref{fig:mainplot_type1}, but here $104$ additional
snapshots from the $4$ additional runs are considered: this resulted
in an increase of the maximum amount of mass segregation seen in
simulations, still none of the snapshots fully reaches the data
points.}
\label{fig:mainplot_type2}
\end{figure}

We have addressed possible sources of systematic errors in
constructing the observed mass segregation profile, namely
mis-estimation of the half-mass radius $r_{hm}$ and cluster
miscentering (see Section~\ref{sec:syst}). Neither of these is expected to
have a significant impact on our measure. Quantifying the robustness
of the measured run-to-run variations from the sample of our
simulations is instead more challenging. While the original
\cite{gill08} simulations have $162$ snapshots with an IMBH and
it is thus hard to solidly assess fluctuations above the
$2 \sigma$ level, Fig. \ref{fig:mainplot_type2} shows that
the inclusion of $104$ more snapshots from runs with an IMBH
does not alter our main results, thereby showing that both
systematic and statistic errors are under control.

Despite the null result of this search, we have demonstrated the
practical application of mass segregation as a fingerprint for the
presence or absence of an IMBH. The method can be applied to other
galactic globular clusters provided that data of HST quality for the
central regions of the systems are complemented by the acquisition of
a field located around the half-light radius of the system.

\acknowledgements

We thank Enrico Vesperini for useful discussions and suggestions and
the referee for a thorough and constructive report. Support for
proposal HST-AR-11284 was provided by NASA through a grant from STScI,
which is operated by AURA, Inc., under NASA contract NAS 5-26555. This
work was supported in part through NASA ATFP grant NNX08AH29G and by
the National Science Foundation under Grant No. PHY05-51164.

\bibliography{ms}{}

\clearpage

\clearpage

\rotate
\begin{deluxetable}{rrrrrr|rrrrrr}
\tablecolumns{12} 
\tablewidth{0pc} 
\label{tab:sim}
\tablecaption{Summary of the N-body simulations.}
\tablehead{\colhead{ID} &\colhead{N} & \colhead{IMF} & \colhead{$M_{IMBH}$/$M_{tot}$} &
\colhead{$M_{IMBH}$/$M_{\odot}$} & \colhead{$f_b$} & \colhead{$ {\langle \Delta m \rangle}_{G08}  $}& \colhead{$ {\langle \Delta m \rangle}_{P09} $} & \colhead{$ {\sigma}_{P09} $} & \colhead{${\Delta m}^{\mathtt{min}}_{P09}$} & \colhead{${\Delta m}^{\mathtt{max}}_{P09}$}}
\startdata
16ks    & $16384$ & Sal  & $N/A   $& $N/A   $&$ 0 $ & $0.11$ & $0.101$ & $0.008$ & $0.083$ & $0.113$\\
16ks.1  & $16384$ & Sal  & $N/A   $& $N/A   $&$ 0 $ & $0.14$ & $0.13$  & $0.01 $& $0.112$& $0.158$\\
16km    & $16384$ & M\&S & $N/A   $& $N/A   $&$ 0 $ & $0.14$ & $0.137$ & $0.009$ & $0.116$ & $0.151$\\
16kbs   & $16384$ & Sal  & $N/A   $& $N/A   $&$0.1$ & $0.09$ & $0.074$ & $0.008$ & $0.056$ & $0.090$\\
16kbm   & $16384$ & M\&S & $N/A   $& $N/A   $&$0.1$ & $0.10$ & $0.102$ & $0.009$ & $0.085$ & $0.124$\\
32km    & $32768$ & M\&S & $N/A   $& $N/A   $&$ 0 $&  $0.14$ & $0.142$ & $0.007$ & $0.128$ & $0.154$\\
\tableline
16ksbh  & $16385$ & Sal  & $0.015 $& $103.1 $&$ 0 $ & $0.05$ & $0.048$ & $0.006$ & $0.037$ & $0.060$\\
16ksbh.1& $16385$ & Sal  & $0.015 $& $60.9  $&$ 0 $ & $0.06$ & $0.06$  & $0.01 $ & $0.041$ & $0.078$\\
16kmbh  & $16385$ & M\&S & $0.015 $& $128.2 $&$ 0 $ & $0.08$ & $0.071$  & $0.009 $ & $0.048$ & $0.091$\\
16kbsbh & $16385$ & Sal  & $0.01  $& $113.4 $&$0.1$ & $0.04$ & $0.042$ & $0.005$ & $0.033$ & $0.052$\\
16kbmbh & $16385$ & M\&S & $0.01  $& $141.0 $&$0.1$ & $0.05$ & $0.050$ & $0.008$ &  $0.038$ & $0.072$\\
32kmbh  & $32769$ & M\&S & $0.01  $& $240.0 $&$ 0 $ & $0.07$ & $0.069$ & $0.005$ & $0.058$ & $0.083$\\
\tableline
\tableline
16kmbh$^{I}$  & $16385$ & M\&S & $0.015 $& $128.2 $&$ 0 $ & $N/A$ & $0.069$  & $0.008 $&$0.055$& $0.083$\\
16kmbh$^{II}$  & $16385$ & M\&S & $0.015 $& $128.2 $&$ 0 $ & $N/A$ & $0.063$  & $0.010 $&$0.040$ & $0.079$\\
16kmbh$^{III}$  & $16385$ & M\&S & $0.015 $& $128.2 $&$ 0 $ & $N/A$ & $0.055$  & $0.009 $&$0.042$ & $0.074$\\
16kmbh$^{IV}$  & $16385$ & M\&S & $0.015 $& $128.2 $&$ 0 $ & $N/A$ & $0.074$  & $0.009 $&$0.060$ & $0.093$\\
\enddata
\tablecomments{Properties of the N-body simulations used in this
paper. The first column is the simulation ID (subscript .1 means that
the IMF was down to $0.1 \mathrm{M_{\sun}}$), the second column
reports the number of particles in the run, the third the IMF
(\citealt{salpeter} or \citealt{millerscalo}), the fourth the BH to
total mass ratio, the fifth the BH mass in solar units, the sixth the
primordial binary fraction. The seventh entry, ${\langle \Delta m
\rangle}_{G08}$, is the snapshot-time-averaged value for the
difference between mean main-sequence mass at the center of the
cluster and mean main-sequence mass around $r_{hm}$, according to the
definition of \cite{gill08}.  The eight entry, ${\langle \Delta m
\rangle}_{P09}$, is the same quantity but normalized as discussed in
section~2.2. Its standard deviation, minimum and maximum values are
given in the last three columns. Snapshots from $t=7 t_{rh}$ to
$9~t_{rh}$ are used in the analysis. The first 12 entries are the
simulations discussed in \cite{gill08}. The last 4 entries are
additional runs, that are ``randomized'' clones of 16kmbh (generated
with a different random number seed).}
\end{deluxetable}

\clearpage

\end{document}